# Field demonstration of a fully managed, L1 encrypted 3-node network with hybrid relayed-QKD and centralized symmetric classical key management

**N. Makris [1], A. Papageorgopoulos [1], K. Tsimvrakidis [1], P. Konteli [1], Y. Gautier[2], M. Terenziani[2], E. Daudin[2], D. Ntoulias[2], T. Fragkioudakis[2], I. Meletios[2], M. Mosca [3], D. Hobbs[3], T. Rosati[3], I. Papastamatiou[4], O. Prnjat [4], K. Koumantaros [4], D. Mitropoulos [1, 4], Jean-Robert Morax [5], Bruno Huttner [5], O. K. Christodoulopoulos[1], G. T. Kanellos [1,*], D. Syvridis [1]**

[1] National and Kapodistrian University of Athens, Athens, Greece
[2] Nokia, Espoo, Finland
[3] EvolutionQ Inc., Waterloo, Canada
[4] GRNET S.A. – National Infrastructures for Research and Technology, Athens, Greece
[5] ID QUANTIQUE S.A. - Geneva, Switzerland
*gtkanellos@di.uoa.gr

**Abstract:** We successfully demonstrated a fully-managed, field-deployed, three-node QKD ring network with L1-OTNsec encryption, that employs a hybrid scheme of QKD and classical yet quantum-safe centrally-generated symmetric keys to support point-to-point and relay consumers.

## 1. Introduction

Quantum networks leverage quantum key distribution (QKD) to ensure communication security. To efficiently integrate QKD networks into existing infrastructure and be optimally functional, the European and International QKD standards [1]-[4] have proposed a layered framework, comprising of the quantum layer, the key management (KMS) layer and the application layer. This architecture is essential to enable encrypted communication across various applications and users. The first major quantum network was implemented by DARPA that followed a three-layer architecture and adopted a hybrid switched/relayed implementation. Other developments include the SECOQC network, focused on relayed-QKD (trusted repeater prototype) setups, the Tokyo project [5] and the Cambridge quantum network [6]. Recently, China presented a 46-node quantum metropolitan area network [7] connecting 40 user nodes including three trusted relays and three optical switches. However, without centralized orchestration, the management of the network remains suboptimal and inefficient. Software – Defined QKD (SDQKD) offers a potential solution to address this issue and improve the efficiency and flexibility of a network. The Madrid SDQKD was the first successful full-scale integration of QKD technology in an SDN environment serving encrypted communication between 3 relayed nodes [8]. The Subcarrier Wave [9] was utilized to enable communication in a 3 node SDN enabled network configuration. [10] Finally, a recent development involved a new framework for Software Defined Network for QKD as a service (SDQaaS) [10].

In this work we present, to the best of our knowledge, the first fully managed, operational three-node quantum cryptography network serving OTN circuits with (Layer 1) encryption. Fig. 1a illustrates the architecture of our network. Each node implements a fully integrated vertical stack featuring a managed quantum layer, a KMS layer and, an application layer that are all jointly orchestrated. In particular, the quantum layer features two IDQuantique Cerberis XGR QKD pairs used to establish: i) an all-to-all communication, and ii) 3 (bidirectional) point-to-point optical circuits using a relayed QKD configuration. In this way our system minimizes the QKD resources as the QKD pair for the third link is omitted. The quantum layer complies to the ETSI GS QKD 014 for secure quantum key extraction and delivery to the KMS layer [1]. On the KMS layer, we deployed an EvolutionQ BasejumpQDN key management software and the Network Controller for the QKDN that acts as per ETSI GS QKD 015, and pulls the generated quantum keys, stores them in dedicated buffers and synchronizes and schedules the key delivery to the Secure Application Entities (SAEs). Note that the network encrypts OTN circuits between all the 3 nodes pairs while employing only 2 QKD pairs, allowing the key consumption in the intermediate relay node while enabling the application of priorities or QoS attributes to the key consumption. Such an addition is most valuable, but the key management process requires additional scheduling, buffering and key monitoring. The KMS layer is controlled and managed by the Nokia 1830 Security Management Server (SMS) fitted to control the EvolutionQ key management software. The SMS functions as a security network orchestrator, similar to the ETSI GS QKD 18. Specifically, it orchestrates the keys across the nodes of the network but also coordinates the key requirements with the classical OTN layer. Additionally, in a novel feature not demonstrated before in quantum encrypted networks, the SMS offers seamless transition to a classical (yet quantum-safe) encryption in the event of a DDOS (Distributed Denial of Service) attack on the QKD layer or a QKD failure in key generation/distribution. To do so, it implements a scheme of centralized symmetric classical keys distribution and encryption based on AES256. On the application layer, each node hosts a Nokia layer 1/OTN Photonic Service Interconnection - Modular (PSI-M) encryptor (300Gbps) for encrypting data with the requested quantum keys at key rotation rates as small as 1key/min.

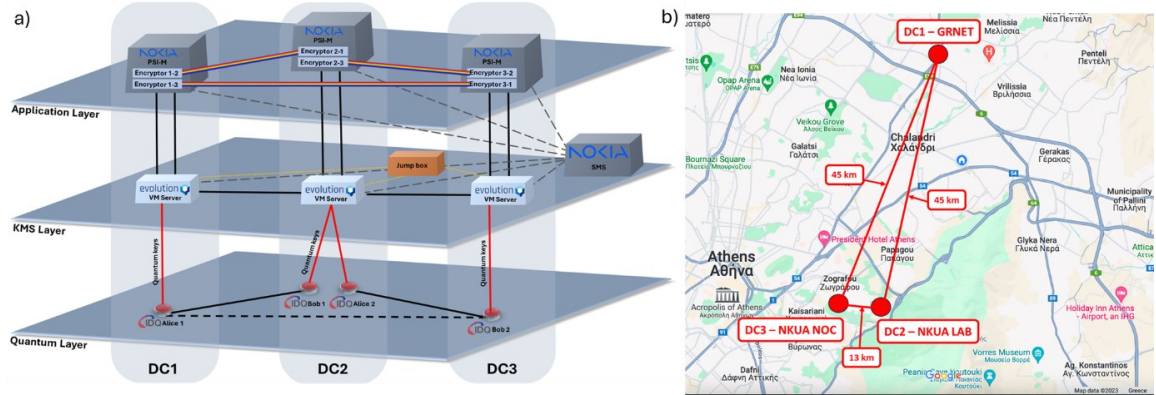

Fig. 1: a) Illustration of the layered architecture depicting the quantum layer, the KMS layer, the application layer and their interconnections. b) DC1, DC2 and DC3 marked by their approximated physical location on the Athens city map.

## 2. Experimental Setup

The experimental testbed, shown in Fig. 1b, comprise 3 trusted nodes: DC1 (GRNET DC node), DC2 (NKUA Optical Communications and Photonic Technologies Lab) and DC3 (NKUA Networks Operation Center-NOC). DC1 hosts the QKD Alice1 (A1), DC2 is the relay node and hosts QKD Bob1 (B1) and Alice2 (A2), and DC3 hosts QKD Bob2 (B2). The fiber link between DC1 and DC2 is ~45km, while DC2 and DC3 are in separate sites in NKUA campus and connected through a fiber spool of ~13km to emulate a longer distance.

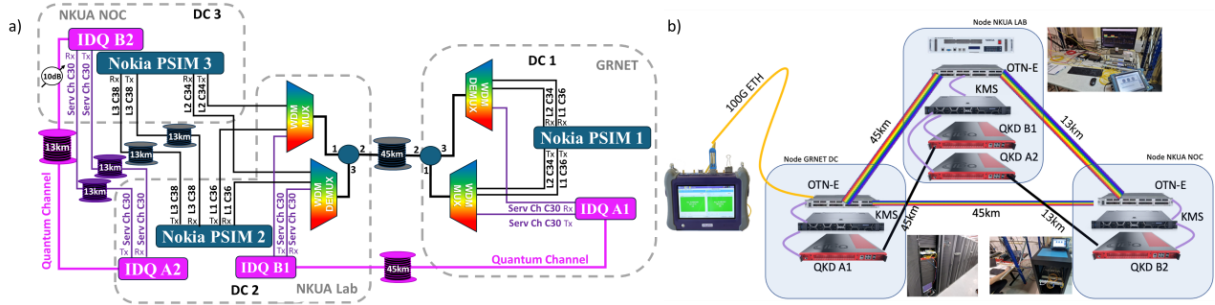

Fig. 2: a) Optical connections of the three-node relayed quantum network. b) Full stack schematic demonstration and its real counterpart.

The physical optical connectivity for the QKD pairs and the PSI-M encryptors are demonstrated in Fig. 2. DC2 and DC3 (NKUA nodes) are connected to DC1 (GRNET) through two unidirectional dark fibers. The 1550nm quantum channel occupies one of the two dark fibers to avoid in-band noise that would degrade the QKD performance, deemed essential as the A1-B1 link encounters losses of 18dB, approaching the operational limit for the IDQuantique Cerberis XGR QKD pair. The PSI-M OTN data channels (C36, C34) and the QKD service channel (C30) are multiplexed together, and circulators were employed to operate the second fiber in both directions. The additional losses were 2.8dB in total. The OTN data channel between the PSI-M 2 and PSI-M 3 (C38 channel) is established via an intra-campus link also used for the communication between Alice 2 and Bob 2. Every PSI-M line has a configuration that supports a data rate of 300Gbps via 67GBaud and 16SQAM with SDFEC-G2 for flex OTU in OTSiG. Regarding the management network, each node has a private ethernet connection between the QKD and KMS server to deliver the secure generated quantum keys from the quantum layer to the KMS layer. Moreover, a VLAN in the GRNET-NKUA domain was established to facilitate management, orchestration, and communication among the components within the deployed multi-layer architecture.

## 3. Results and Discussion

The system was tested under different key rotation intervals (1, 5, 15, 60 min) for extended periods. Table 1 includes the percentages for the various key rotation outcomes. We observed 98% successful key rotation using QKD or classical key distribution in the stress condition of 1key/min. Network redundancy was tested by emulating two separate attacks: i) a KMS attack causing a temporary disruption in the communication between the application and KMS layers and ii) a QKD attack that would nullify SKR. The KMS attack is emulated by shutting down the KMS server in each node, whereas the QKD attack is emulated by plugging off the quantum channel fiber. For both tests the key rotation interval was set to 1 minute, the QNL buffer (EvolutionQ KMS input buffer where keys are stored after retrieved from the quantum layer) capacity was 1000 keys, and the key expiration time was set to 4 hours. The results are demonstrated in Fig. 3.

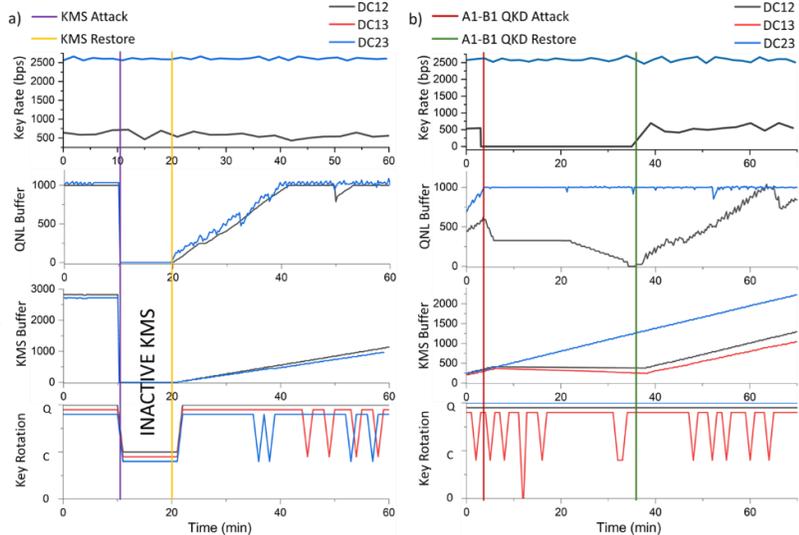

Table 1: Network's key rotation percentages for different key rotation intervals.

| Rotation Interval (min) | Classical Key Rotation (%) | Quantum Key Rotation (%) | SNMP error (%) |
|---|---|---|---|
| 1 | 8.8 | 89.2 | 2.0 |
| 15 | 14.8 | 83.6 | 1.6 |
| 30 | 33 | 66,4 | 0,6 |
| 60 | 7.8 | 90,7 | 1.5 |

Table 2: Network Analyzer Average Statistics

| Throughput (Gbit/s) | Jitter (ms) | Latency (ms) | Frame Loss |
|---|---|---|---|
| 158.4 | 8.8 | 89.2 | 0.0 |

Fig. 3: Joint diagrams for the SKR, QNL buffers, KMS buffers and SMS key rotation for all links at the emulated KMS (a) and QKD (b) attacks. The key rotation is characterized as 'Q' for quantum key rotation, 'C' for the SMS' classical key rotation and, '0' for no key rotation.

As presented in Fig. 3a, by emulating a KMS attack on the network, its hybrid operation is demonstrated. By launching a KMS attack in each server, the KMS layer cannot provide the encryptors with quantum keys. In that case, the SMS generates and provides the PSI-Ms with classical quantum-safe keys. Once the servers become available again, the QNL (KMS input) buffers start to fill up again. After a brief interval, they begin supplying the KMS (output) buffers with keys which are then used for quantum key rotation. Accordingly, in Fig. 3b the QKD attack is demonstrated in DC12. When the attack occurs, we use the buffered keys. There is a linear reduction of keys (plus some overhead). Then depending on the attack duration, we can run out of keys or avoid running out of keys. In the experiment the duration was short enough and we managed to get away with no interruption of quantum keys. Once the QKD link was restored, the QNL buffers began to replenish, consequently extending the buffering process within the KMS.

### 4. Conclusion

We report on the successful demonstration of a fully managed field deployed three-node QKD relayed network in a ring configuration, where each node comprises a fully integrated vertical stack. We demonstrated the network stable operation which provides data encryption in the physical layer with quantum keys. In case of an attack or failure where quantum keys are unavailable, we demonstrated the capability of seamlessly transition to classically encrypted quantum-safe keys provided by the central security manager / SMS, exhibiting, thus, a reliable hybrid network operation and ensuring the continuous supply of keys. Notably, the keys were consumed in all three nodes, including the intermediate relayed node, and we demonstrated that the operations, including relaying, recovers against various attacks.

### 5. Acknowledgements

This work was funded by the EU quantum flagship project QSNP (GA 101114043) and the HellasQCI project (GA 101091504).